\documentclass[
               twocolumn,
               noshowpacs,          
               nopreprintnumbers,     
               aps,                 
               prd,                 
               a4paper,             
               superscriptaddress,  
               nofootinbib,         
               tightenlines,        
               floats,floatfix      
               ]{revtex4}

\usepackage{amsmath, amsfonts, amsthm, amssymb, graphicx}
\usepackage{epstopdf}
 \usepackage{slashed}
 \usepackage{graphicx,epsfig}
\usepackage{amsmath}
\usepackage {amssymb}

\newcommand{\be}{\begin{eqnarray}}
\newcommand{\ee}{\end{eqnarray}}
\newcommand{\bea}{\begin{eqnarray}}
\newcommand{\nn}{\nonumber}
\newcommand{\eea}{\end{eqnarray}}

\def\a{\alpha}
\def\b{\beta}
\def\g{\gamma}

\def\d{\delta}

\def\La{\Lambda}
\def\k{\kappa}
\def\m{\mu}
\def\n{\nu}

\def\t{\tau}

\def\z{\zeta}

\begin{document}

\title{Warped-AdS$_3$ black holes with scalar halo}

\author{Gaston Giribet}
\email{gaston-at-df.uba.ar}
\affiliation{Physique Th\'eorique et Math\'ematique, Universit\'e Libre de Bruxelles and International Solvay Institutes,
Campus Plaine C.P. 231, B-1050 Bruxelles, Belgium.}
\affiliation{Departamento de F\'{\i}sica, Universidad de Buenos Aires and IFIBA-CONICET,
Ciudad Universitaria, Pabell\'on I, 1428, Buenos Aires, Argentina.}
\affiliation{Instituto de F\'{\i}sica, Pontificia Universidad Cat\'{o}lica de Valpara%
\'{\i}so, Casilla 4059, Valpara\'{\i}so, Chile.}
\author{Minas Tsoukalas}
\email{minasts-at-cecs.cl}
\affiliation{Centro de Estudios Cient\'{\i}ficos CECs, Casilla 1469, Valdivia, Chile.}
\pacs{3.141592653589793238462643383...}

\date{\today}

\begin{abstract}
We construct a stretched (aka Warped) Anti-de Sitter black hole in $3$ dimensions supported by a real scalar field configuration. The latter is regular everywhere outside and on the horizon. No-hair theorems in $3$ dimensions demand the matter to be coupled to the curvature in a non-minimal way; however, this coupling can still be of the Horndeski type, i.e. yielding second order field equations similar to those appearing in the context of Galileon theories. These Warped-Anti-de Sitter black holes exhibit interesting thermodynamical properties, such as finite Hawking temperature and entropy. We compute the black hole entropy in the gravity theory and speculate with the possibility of this to admit a microscopic description in terms of a dual (Warped) Conformal Field Theory. We also discuss the inner and outer black hole mechanics.
\end{abstract}

\maketitle

\section{Introduction} 

One of the most interesting proposals to extend AdS/CFT holographic duality to non-AdS scenarios is the so-called WAdS/CFT correspondence, 
which involves a specific warped deformation of the $3$-dimensional Anti-de Sittrer space. WAdS/CFT correspondence was originally conjectured in Ref. \cite{Strominger}, where it 
was proposed that quantum gravity about $3$-dimensional Warped Anti-de Sitter space (WAdS$_3$) is dual to a $2$-dimensional conformal field theory (CFT$_2$). It 
was later realized in Ref. \cite{DHH} that the two-dimensional theory that is supposed to be dual to WAdS$_3$ gravity could be of a different sort, exhibiting a 
special type of conformal symmetry generated by only one copy of Virasoro algebra and one copy of the affine $\hat{u}(1)$ Kac-Moody algebra \cite{a1}-\cite{a4}; see \cite{Geoffrey} for an alternative proposal.

In the last years, WAdS$_3$ spaces and WAdS/CFT correspondence were intensively studied in different contexts \cite{a1}-\cite{HofmanRolliest}, including massive gravity, string 
theory, higher-derivative models, and topologically massive gauge theories. Besides, these spaces are also connected with other holographic realizations, like the Kerr/CFT correspondence 
\cite{preKerrCFT,KerrCFT}, and cond-matt/CFT \cite{Son, BalasubramanianMcGreevy}.

One of the salient features of WAdS$_3$ space is the existence of black hole geometries that, while being asymptotically WAdS$_3$, are locally equivalent to 
WAdS$_3$ itself at any point. This is similar to what happens in AdS$_3$ space, where the Ba\~{n}ados-Teitelboim-Zanelli (BTZ) black holes \cite{BTZ} are asymptotically 
AdS$_3$ and, at the same time, are locally equivalent to AdS$_3$.

The asymptotically WAdS$_3$ black holes (hereafter referred to as WAdS$_3$ black holes) were first derived in Ref. \cite{bha} (see also \cite{Strominger, bhz}) and were crucial ingredients to formulate the 
holographic conjectures of \cite{Strominger} and \cite{DHH}. These solutions were first found in the context of topologically massive gauge theories, whose actions correspond to augmenting the $(2+1)$-dimensional Einstein-Maxwell action with a Chern-Simons term either for the affine connection and/or for the gauge connection. Later, these solutions were also found in theories with no Chern-Simons terms, such as $R^2$-gravity models. In this paper, we raise the question as to whether WAdS$_3$ black holes exist in much simpler models, e.g. whether they can be supported by field configuration in $(2+1)$-dimensional Einstein theory with neither Chern-Simons terms nor higher-curvature terms. This question has been addressed before, for instance in Ref. \cite{anninos}; however, the model considered here is of a different kind: We are interested in finding a minimal sensible scenario in which WAdS$_3$ black holes can be supported by a real scalar field. That is, we consider $(2+1)$-dimensional Einstein theory coupled to scalar field matter. Of course, the first obstruction that appears is that no-hair theorems in $3$ dimensions demand the scalar field matter to be coupled to the curvature in a non-minimal way. A priori this could seem to imply that exotic matter couplings are needed. However, we show here that, still, WAdS$_3$ black hole solutions are possible in scalar-tensor models with second order field equations. The model we consider is the $3$-dimensional analog of the so-called Horndeski's theory, namely the most general scalar-tensor theory yielding second order field equations in $4$ dimensions \cite{Horndeski:1974wa}. This theory is similar to the Galileon models \cite{Galileons}; see also \cite{Deffayet:2011gz} and references therein. Apart from the usual kinetic term $\sim g^{\mu \nu} \nabla_{\mu}\phi \nabla_{\nu}\phi$ for the scalar field $\phi $, the Lagrangian involves a term $\sim G^{\mu \nu} \nabla_{\mu}\phi \nabla_{\nu}\phi$, with $G_{\mu \nu}=R_{\mu \nu}-(1/2)Rg_{\mu \nu}$ being the Einstein tensor. These type of couplings arises naturally through dimensional reduction \cite{VanAcoleyen:2011mj} and have been recently studied in several context; see for instance \cite{fab4, Charmousis:2014mia} and references therein and thereof.

We will introduce the scalar-tensor theory in section II. In section III, we will study they locally AdS$_3$ solutions and a specific type of deformations of them. In section IV, we will discuss the WAdS$_3$ space and WAdS$_3$ black holes. The thermodynamics of the latter will be discussed in section V, where we also discuss the possibility of a holographic dual description.

\section{Scalar-tensor theory}

Consider the theory defined by the action
\bea\label{action}
&& {\mathcal I}=\int dx^{3} \sqrt{-g} \Big( \z R-2\La\nn\\
&& \qquad\qquad-\eta g^{\m\n}\nabla_{\m}\phi\nabla_{\n}\phi+\b G^{\m\n}\nabla_{\m}\phi\nabla_{\n}\phi\Big) ,
\eea
where $\beta $ is a real parameter with mass dimension $-2$, while $\eta $ is a parameter that, without loss of generality, can be considered to take values $\eta =0, \pm 1$. Having set the Planck scale to $1$, parameter $\zeta =\pm 1$ stands for controlling the sign of the effective Newton constant. 

The scalar field in (\ref{action}) experiences an effective geometry with metric $\bar{g}_{\mu \nu } = g_{\mu \nu } -(\beta/\eta) G_{\mu \nu }$, which in the case we will be interested in turns out not to be conformally flat. 

The field equations coming from varying the action with respect to the metric $g_{\mu \nu }$ and the scalar field $\phi $ are
\bea\label{metriceq}
\mathcal{E}_{\m\n}&=&\z G_{\m\n}-\eta \left(\nabla_{\m}\phi\nabla_{\n}\phi-\frac{1}{2}g_{\m\n}\nabla^{\a}\phi\nabla_{\a}\phi \right) \nn\\
&&+g_{\m\n} \La+\frac{\beta}{2} \Big((\nabla^{\a}\phi \nabla_{\a}\phi)\, G_{\m\n}\nn\\
&&+g_{\m\a} \d^{\a\rho\sigma}_{\n\g\d}\nabla^{\gamma}\nabla_{\rho}\phi\nabla^{\d}\nabla_{\sigma}\phi \Big)=0,
\eea
and
\be\label{scaleq}
\nabla_{\m}J^{\m}=0, \ \ \ J^{\m}=(\eta g^{\m\n}-\beta G^{\m\n})\nabla_{\n}\phi ,
\ee
respectively. 

In $4$ dimensions, the analogous field equations contain an additional term that involves the double dual of the Riemann tensor, which in $3$ dimensions vanishes identically. Black hole solutions with non-trivial derivative couplings of the type (\ref{action}) in $4$ dimensions have been recently studied in Refs. \cite{Kolyvaris:2011fk}-\cite{Bravo-Gaete:2013dca}, some of them accomplishing to circumvent no-hair arguments \cite{Hui:2012qt}-\cite{Sotiriou:2013qea}. Extensions to the multi-scalar case have also been investigated in \cite{Charmousis:2014zaa}, while solutions in higher dimensions including higher curvature derivative couplings, which still maintain second order field equations where found in \cite{Charmousis:2015txa}. In $3$-dimensions, this type of couplings was considered in Ref. \cite{Bravo-Gaete:2014haa}, where the asymptotically AdS$_3$ black holes were studied. Here, we extend the results of \cite{Bravo-Gaete:2014haa} to asymptotically WAdS$_3$ black holes. 

\section{AdS$_3$ waves and black holes}

Let us begin by studying the simplest type of deformation of locally AdS$_3$ spaces: the so-called AdS-waves. This amounts to considering the ansatz
 \be
 ds^{2}=\frac{l^{2}}{r^{2}}\big( -(1+2 h(r))dt^{2}+2 dt d\xi+dr^{2}\big) \label{crucial}
 \ee
and assuming the scalar field to depend only on $r$. 

With these assumptions, from (\ref{scaleq}) one finds 
\be
(\b-l^{2}\eta) (-\phi'(r)+r \phi''(r))=0, \label{crucia}
\ee
where the prime stands for derivatives with respect to $r$. 

One sees from (\ref{crucia}) that there exist two distinct branches of solutions: the one that makes the first factor in (\ref{crucia}) to vanish, and the one that makes the second factor in (\ref{crucia}) to vanish. Let us consider these two branches separately in the following two subsections.

\subsection{The quadratic branch} \label{case1}

First, let us examine the case where the second factor in (\ref{crucia}) vanishes. Solving for $\phi$, one finds the following quadratic form
\bea
\phi(r)=\frac{1}{2} c_{1} r^{2}+c_{2},
\eea 
where $c_{1}$ and $c_{2}$ are integration constants. Substituting this into (\ref{metriceq}), one finds the following two constraints
\be
\frac{3 \b}{l^{2}}-\eta=0\,\,\,\text{and}\,\,\,\z+l^{2}\La=0.
\ee

Then, one is left with only one equation; namely
\be
(3c_{1}^{2}r^{4}\b+2l^{4}\La)h'(r)+r(c_{1}^{2}r^{4}\b-2l^{4}\La)h''(r)=0.
\ee
Solving this equation, one gets the kink-type profile
\be
h(r)=-h_1 {\text{arctanh} \left( \frac{c_{1}r^{2}}{l^{2}}\sqrt{\frac{\b}{2\La}}\right) }+h_{2}, \label{19}
\ee
where $h_{1}$ and $h_{2}$ are integration constants. The case $h_1=0$ yields a locally AdS$_3$ solution; in particular, this leads to the BTZ black hole solution studied in Ref. \cite{Bravo-Gaete:2014haa}. 

While the hyperbolic tangent in (\ref{19}) corresponds to the case in which $\text{sign}(\beta ) = \text{sign}(\Lambda )$, the solution can be analytically continued to the case $\text{sign}(\beta ) = -\text{sign}(\Lambda )$. In the former case and for negative $\La$ one finds that $\beta$ is also negative and the scalar field has the wrong sign in its kinetic term. On the other hand, if $\text{sign}(\beta ) = \text{sign}(\Lambda )$ and $\La $ is positive, then the scalar field kinetic has the correct sign, but the Einstein-Hilbert term switches its sign. In the latter case, i.e. when $\text{sign}(\beta ) = -\text{sign}(\Lambda )$, $\La $ can be negative for positive values of both $\beta $ and $\z $.

\subsection{The logarithmic branch} \label{case2}

The second branch corresponds to having
\be
\b-l^{2}\eta=0.
\ee

Then, from the $\mathcal{E}_{rr}=0$ equation one finds
\bea
\phi(r)=\pm\sqrt{\frac{-l^{2}(\z+l^{2}\La)}{\beta}}\log (r)+\phi_{1},
\eea
with $\phi_{1}$ being an integration constant. Substituting this into the $\mathcal{E}_{tt}=0$ equation, for $\z\neq l^{2}\La$, one gets
\bea
h(r)=\frac{1}{2}r^{2}h_{1}+h_{2}.
\ee

In the case $\La = \z/l^{2}$, on the other hand, the metric function $h$ is left undetermined \cite{Bravo-Gaete:2014haa}. If $\La>\z/l^{2}$ then $\b<0$ and $\eta<0$, while if $\La<\z/l^{2}$ then $\b>0$ and $\eta >0$. The latter case yields the correct sign for the kinetic term.

\section{WAdS$_{3}$ space and black holes}

\subsection{WAdS$_3$ space}

Now, let us move to consider the asymptotically WAdS$_3$ solutions. WAdS$_3$ space is a specific deformation of AdS$_3$ space that consists of 
squashing or stretching AdS$_3$, in a similar manner as one uses to stretch or squash the $3$-sphere \cite{preKerrCFT}. Following \cite{Strominger}, one can represent WAdS$_3$ space by first writing AdS$_3$ as a Hopf fibration of the real line over 
AdS$_2$, and then multiplying the fiber by a warp constant factor $K$. In this way, one obtains a geometry whose metric reads
\be\label{wadsmetric}
ds^{2}=\frac{l_{K}^{2}}{4}\Big(-\cosh^{2}x \,d\t^{2}+dx^{2}+K (dy+\sinh{x}\,d\t)^{2} \Big)
\ee 
where $K$ and $l_K$ are two real parameters. Provided coordinates $x$, $y$ and $\t $ take values on the real line, the case $K=1$ of (\ref{wadsmetric}) reduces to AdS$_3$ space; while for $K\neq 1$ it represents a non-Einstein space with isommetry group $SL(2,R)\times U(1)$. 

It is convenient to introduce two new parameters $\nu $ and $l$, defined by 
\be
K=\frac{4\nu^{2}}{(\nu^{2}+3)} , \ \ \ \ \ l^{2}_{K}=\frac{4l^{2}}{(\nu^{2}+3)} .
\ee
Locally AdS$_3$ spaces then correspond to $\nu^2=1$. Spaces with $\nu^2>1$ ($\nu^2<1$) represent stretched (resp. squashed) deformations of AdS$_3$. 

The scalar curvature associated to (\ref{wadsmetric}) is
\be
R=\frac{2 (K-4)}{l^{2}_{K}} = - \frac{6}{l^{2}}\nn ,
\ee
which is always negative, provided $\nu^2 >0$.

Assuming the scalar field to be a function of the coordinate $x$, its field equation yields
\be
(K\,\b-\eta\,l_{K}^{2}) \left(\tanh{x}\, \phi'(x)+\phi''(x) \right)=0 , \label{dos}
\ee
where the prime now stands for the derivative with respect to $x$. 

As in the AdS$_3$ case discussed in the previous section, there are two branches to consider: The analogous to the quadratic branch studied in subsection \ref{case1} corresponds to considering the second factor in (\ref{dos}) to be to zero. In that case, the scalar field can be easily integrated and shown to give
\be
\phi(x)=\phi_0 \ \text{arctanh} ( e^x ) + \phi_1 \label{sfc}
\ee
where $\phi_0$ and $\phi_1$ are arbitrary real constants. However, one can easily check that, for the case of WAdS$_3$ solutions, (\ref{sfc}) is not consistent with the rest of the field equations. This is actually expected, as (\ref{sfc}) tends to a constant when $x$ goes to infinity. Thus, one has to consider the other branch, namely 
\be\label{wadsconst1}
K\,\b-\eta\,l_{K}^{2}=0, \label{eeesta}
\ee
which is analogous to the logarithmic branch studied in subsection \ref{case2} for the AdS$_3$ case.

Eq. (\ref{eeesta}) implies
\be\label{wadsconst1-2}
\nu^2=\frac{l^2\eta }{\beta },
\ee
which in particular says that $\text{sign}(\beta)=\text{sign}(\eta)$.

From the $\mathcal{E}_{xx}=0$ equation, one has 
\bea 
\phi (x) = \pm \sqrt{-\frac{K\zeta +l_K^2\Lambda}{4\eta }} x + \phi_1 \nn  \label{wadsscalar}
\eea 
where $\phi_{1}$ is an integration constant. Substituting this into (\ref{metriceq}) one finds 
\be \label{wadsconst2}
K\,\z-l^{2}_{K}\,\La=0. 
\ee
which implies
\be \label{wadsconst2e}
\nu^2 =\frac{l^{2}\Lambda }{\zeta } =\frac{l^2\eta }{\beta } . 
\ee

Then, the solution for the scalar field can be written as
\be
\phi (x)=\pm\sqrt{-\frac{2\nu^{2}\zeta / \eta }{(\nu^{2}+3)}}x+\phi_{1} . \label{20}
\ee

For positive $\z$ one needs to have $\b<-{l^{2}\eta}/{3}$, which translates into $(3+\nu^{2})\b<0$. Negative $\beta$ implies having the wrong sign for the kinetic piece. On the other hand, when $\z$ is negative then  $\b>-{l^{2}\eta}/{3}$ and so $(3+\nu^{2})\b>0$, yielding the correct sign for the kinetic piece of the scalar. However, in the latter case, one finds the opposite sign for the Newton constant.

\subsection{WAdS$_3$ black holes}

We have just seen that the WAdS$_{3}$ space (\ref{wadsmetric}) solves the field equations (\ref{metriceq})-(\ref{scaleq}). Therefore, the same can be shown to happen for WAdS$_{3}$ black holes, provided they are locally equivalent to the former. Let us see this explicitly: In its Arnowitt-Deser-Misner form, the WAdS$_{3}$ black hole metric reads
\bea\label{WAdS3BH}
&&ds^{2}=-N(r)^{2}dt^{2}+l^{2}\,R(r)^{2}\left(d\theta+N^{\theta}(r)dt \right)^{2}\nn\\
&&\qquad+\frac{l^{4}}{4 R(r)^{2}\,N(r)^{2}}dr^{2},
\ee
with metric functions
\bea
N(r)^{2}&=&\frac{l^{2}\,(\nu^{2}+3)\,(r-r_{+})\,(r-r_{-})}{4\,R(r)^{2}} \label{N}\\
N^{\theta}(r)&=&\frac{2\,\nu \,r-\sqrt{r_{+}\,r_{-}\,(\nu^{2}+3)}}{2\,R(r)^{2}} \label{Ntheta}\\
R(r)^{2}&=&\frac{r}{4}\,\Big(3(\nu^{2}-1)r+(\nu^{2}+3)(r_{+}+r_{-})\nn\\
&&\qquad \qquad\qquad-4\nu \,\sqrt{r_{+}r_{-}\,(\nu^{2}+3)} \Big),\label{R}
\eea
and with $t\in R$, $r\in R_{\geq 0}$ and $l\theta =[0,2\pi)$. The additional condition $\nu^2>1$ is required for the exterior solution not to exhibit closed timelike curves. In (\ref{N})-(\ref{R}), $r_{+}$ and $r_-$ are two integration constants, which, provided $r_{+}\geq r_{-} \geq 0$, can be interpreted as the outer and inner horizons of the black hole, respectively. At large $r$, black hole metric (\ref{WAdS3BH})-(\ref{R}) asymptotes stretched WAdS$_3$ space. In this limit, one can identify $x\sim \log(r)$.

Assuming $\phi $ only depends on $r$, one finds
\be\label{phibhscal}
\phi(r)=\pm \phi_{0}\,\log(\sqrt{r-r_{+}}+\sqrt{r-r_{-}})+\phi_{1},
\ee
with
\be\label{phibhscal0}
\phi_{0}=2\sqrt{\frac{2\,(l^{2}\,\nu^{2}\,\z+l^{4}\,\La)}{(3+\nu^{2})\,(l^{2}\,\eta-3\nu^{2}\,\b)}},
\ee
and with the constraints
\bea
2\,\nu^{2}\,\b\,\z-l^{2}\,(\z\,\eta+\b\,\La)&=&0  , \label{const1}\\
\nu^{2}\z-l^{2}\La&=&0.\label{const2}
\eea

The scalar field can be equivalently written as
\be
\phi(r)=\pm2\sqrt{-\frac{2\nu^2\z / \eta}{(\nu^2+3)}}\, \log(\sqrt{r-r_{+}}+\sqrt{r-r_{-}})  \label{29}
\ee
plus an arbitrary constant, where it has been used that (\ref{const1}) and (\ref{const2}) implies $\nu^{2}\b-l^{2}\eta=0$. One verifies that, indeed, (\ref{29}) tends to (\ref{20}) as $r \sim e^x$ for large $r$. Again, the reality condition for the scalar depends on the same parameter space as in the case of the WAdS$_{3}$ space.  

Notice that field configuration (\ref{29}) is regular everywhere outside and on the horizon for solutions with $r_+>r_-$. In the extremal case $r_-=r_+$, in contrast, divergence occurs at $r=r_+=r_-$. One can also calculate the energy-momentum tensor 
\bea
&&T_{\m\n}=\eta \left(\nabla_{\m}\phi\nabla_{\n}\phi+\frac{1}{2}g_{\m\n}\nabla^{\a}\phi\nabla_{\a}\phi \right) \nn\\
&&\,\,\,\,\,\,\,\,\,-\frac{\beta}{2} \Big((\nabla^{\a}\phi \nabla_{\a}\phi)\, G_{\m\n}+g_{\m\a} \d^{\a\rho\sigma}_{\n\g\d}\nabla^{\gamma}\nabla_{\rho}\phi\nabla^{\d}\nabla_{\sigma}\phi \Big) \,\,\,\,\,\,\,\,\,\,\,\,
\eea
evaluated on the black hole solution (\ref{WAdS3BH}) and verify that the only non-vanishing components are
\bea
T_{t}^{\,\,\,t}&=&-\frac{(\nu^{2}-3)\z}{l^{2}}\\
T_{r}^{\,\,\,r}&=&\frac{2\nu^{2}\z}{l^{2}}\\
T_{\theta}^{\,\,\,t}&=&\frac{3(\nu^{2}-1)(\sqrt{r_{+}r_{-}(3+\nu^{2})}-2r\nu)\z}{2l^{2}}\\
T_{\theta}^{\,\,\,\theta}&=&\frac{2\nu^{2}\z}{l^{2}} ,
\eea
which are finite at the horizon. The sign of the component $T_{t}^{\,\,\,t}$ depends on whether $\nu^2 >3$ or $\nu^3<0$. The only non-constant component is $T_{\theta}^{\,\,\,t}$, which depends on $r$.

It is worth emphasizing that the $\phi \neq const$ configuration (\ref{29}) does not represent an actual {hair}, as its intensity can not be varied independently of the black hole parameters $r_{\pm }$. This is why we preferred to refer to it as a scalar {halo}, instead.

It is relevant to say that the kinetic term $\sim \dot{\phi }^2$ couples to the effective geometry with a constant factor $-\eta \bar{g}^{tt} = -3\beta (\nu^2-1)/l^4$. Regarding the sign of the action, which is proportional to the volume of the spacetime $\text{Vol}$, it goes like ${\mathcal I}/\text{Vol} = -\zeta l (\nu^2+3)$.

\section{Thermodynamics}

\subsection{Black hole thermodynamics}

Having found WAdS$_3$ black hole solutions in Horndeski's scalar-tensor theory (\ref{action}), one can undertake the study of their thermodynamics. In particular, one can study the black hole entropy $S$ from the gravity point of view and use the result to infer some property of the would-be dual CFT$_2$. 

To compute the entropy, one may resort to Wald's entropy formula, which in $3$ dimensions reads
\be
S_+ = - 2\pi  \oint_{r=r_+} d\theta \sqrt{h}\frac{\d \mathcal{L}}{\d R_{\m\a}}g^{\n\b}\epsilon_{\m\n}\epsilon_{\a\b}. \label{Wald}
\ee
where $\mathcal{L}$ is the Lagrangian density, $\sqrt{h}=\sqrt{g_{\theta\theta}}$ is the area density at the horizon $r=r_+$, and $\epsilon_{\mu\nu}$ are the components of orthogonal bi-vectors. The integration in (\ref{Wald}) is performed on constant-$t$ lines at $r=r_+$. The subindex $+$ in (\ref{Wald}) stands to reminding of the fact that this is the entropy associated to the external horizon (cf. (\ref{Smenos}) below). In the case of (\ref{action}), the integrand gives
\be
\frac{\d \mathcal{L}}{\d R_{\m\a}}=\z g^{\m\a}+\b\nabla^{\m}\phi\nabla^{\a}\phi-\frac{1}{2}\b \nabla^{\k}\phi \nabla_{\k}\phi\, g^{\m\a} ,
\ee
and for the particular case of the WAdS$_{3}$ black holes (\ref{WAdS3BH}) this yields the result
\bea
S_+ = \frac{2\pi}{G} \nu \left( r_+ -\frac{1}{2\nu } \sqrt{(\nu^2+3)r_-r_+} \right) 
\label{S}
\eea
where the Newton constant $G$ has been reintroduced, and it has been used that
\be
\frac{3\nu^{2}\z+l^{2}\La}{\nu}=\frac{4\La \b \nu}{\eta}=4\z \nu . \nn
\ee

Expression (\ref{S}) never vanishes, provided $\nu^2 > 1$ and $r_+\geq r_-$. It does vanish for $r_+ = r_-$ when $\nu^2 = 1$.

The Hawking temperature of WAdS$_3$ black holes, given by their surface gravity, is 
\begin{equation}
T_+ = \frac{(\nu^2+3)}{4\pi l} \frac{(r_+-r_-)}{2\nu r_+ - \sqrt{(\nu^2+3)r_-r_+ }}. \label{TH}
\end{equation}

Another quantity that is relevant for thermodynamics is the angular velocity of the external horizon, which is given by
\begin{equation}
\Omega_+ = \frac{2}{2\nu r_+ - \sqrt{(\nu^2+3)r_-r_+ }} .
\end{equation}

This is the potential associated to the angular momentum in the first principle (see (\ref{1st}) below.)

\subsection{Holography}

At this point, one may feel tempted to follow the ideas of \cite{Strominger} and speculate with the possibility of (\ref{S}) to admit a microscopic description in terms of a dual CFT. Without a complete understanding of the nature of the dual theory to WAdS$_3$ gravity, e.g. knowing whether it is a standard CFT$_2$ or of the type of Warped-CFT$_2$ suggested in \cite{DHH}, one can hardly make a concise statement about such microscopic derivation. However, assuming the existence of Virasoro symmetry at the boundary, one can predict some properties of the boundary theory. 

The WAdS$_3$ black holes are constructed from the global WAdS$_3$ space by means of identifications in two directions. This orbifold type constructions breaks the original $SL(2,R)\times U(1)$ isommetry group to $U(1)\times U(1)$. The inverse of the two identifications periods are
\begin{eqnarray}
&&\beta_L^{-1}= \frac{(\nu^2+3)}{8\pi l^2} (r_++r_--\frac{1}{\nu }\sqrt{(\nu^2+3)r_+r_-})\,\,\,\,\,\,\,\,\,\,\,\, \label{TL} \\
&&\beta_R^{-1}= \frac{(\nu^2+3)}{8\pi l^2} (r_+-r_-) \label{TR}
\end{eqnarray}
and these give the geometrical temperature $T=\beta_L^{-1}+\beta_R^{-1}$, namely
\begin{eqnarray}
T= \frac{(\nu^2+3)}{4\pi l^2} (r_+-\frac{1}{2\nu }\sqrt{(\nu^2+3)r_+r_-}).\,\,\,\,\,\,\,\,\,\,\,\,\, \label{T}
\end{eqnarray}
which has not to be mistaken by the physical temperature (\ref{TH}).

Now, assuming the existence of Virasoro symmetry at the boundary and, as in \cite{Strominger}, invoking the validity of Cardy formula, whose integration version states 
\begin{equation}
S_+ = \frac{\pi^2 l}{3} c T , \label{Cardy}
\end{equation}
one can predict the value of the central charge $c$ of the Virasoro algebra. In this theory, the prediction is
\begin{equation}
c = \frac{24l \nu}{G(\nu^2+3)} , \label{c}
\end{equation}
which tends to infinity in the semiclassical limit $l >> G$.

\subsection{Inner and outer black hole mechanics}

Also motivated by holography, one can perform the analysis of the thermodynamics associated to the internal horizon \cite{Majo}. In fact, at least formally, one can also define thermodynamical quantities analogous to (\ref{S}) and (\ref{TH}) but associated to the horizon located at $r=r_-$. In particular, the entropy associated to this (inner) horizon would be given by integrating Wald's formula on $r=r_-$. This yields
\begin{equation}
S_- = \frac{2\pi \nu }{G} \left( r_- -\frac{1}{2\nu } \sqrt{(\nu^2+3)r_-r_+} \right) . \label{Smenos}
\end{equation}

Analogously, one can define the quantities $T_-$ and $\Omega_-$, associated to the surface $r=r_-$

A first statement of the inner black hole mechanics \cite{Majo} is that the conserved charges, namely the mass ${\mathcal M}$ and the angular momentum ${\mathcal J}$ of the black holes, have to satisfy the first principle of black hole thermodynamics on both horizons; namely
\begin{equation}
d{\mathcal M} = T_{\pm} dS_{\pm} + \Omega_{\pm} \ d{\mathcal J} , \label{1st}
\end{equation}
and this can be used to determine ${\mathcal M}$ and ${\mathcal J}$. Another piece of information comes from the second statement of the inner black hole mechanics; namely the statement that the product of entropies associated to {\it all} the horizons of the solution has to be given entirely by charges that, at quantum level, are expected to be quantized; in this case, the angular momentum ${\mathcal J}$. This follows from consistency of the holographic picture, as the quantization of $S_-S_+$ corresponds to the level matching condition in the dual CFT$_2$. Computing the product of the inner and outer entropies, one finds
\begin{equation}
S_+S_-=  \frac{\pi^2 }{G^2} \left( (5\nu^2+3) r_-r_+ -2\nu \sqrt{(\nu^2+3)r_-r_+} (r_++r_-) \right) ,
\end{equation}
which is precisely the expected result, as it is proportional to the angular momentum of WAdS$_3$ black holes found for other parity-even gravity theories \cite{corea, c2, donnay2}, cf. Ref. \cite{Strominger}. From this argument and from (\ref{1st}) one obtains the following values of the mass and the angular momentum
\begin{equation}
{\mathcal M} = \frac{(\nu^2+3)}{4\nu lG} \left( (r_++r_-)\nu -\sqrt{(\nu^2+3)r_+r_-}\right) \label{masssita}
\end{equation} 
and
\begin{equation}
{\mathcal J} = \frac{(\nu^2+3)}{16\nu lG} \left( (5\nu^2+3)r_-r_+-2\nu \sqrt{(\nu^2+3)r_+r_-} (r_++r_-)\right)  \label{angulito}
\end{equation}
respectively. It can be shown that entropy $S_+$, when written as a function of ${\mathcal M}$ and ${\mathcal J}$, takes a form similar to the Warped version of the Cardy formula proposed in Ref. \cite{DHH}, cf. Ref. \cite{donnay2}. This is consistent with the WAdS/(W)CFT picture.

\section{Conclusions}

In this paper, motivated by the so-called WAdS/(W)CFT correspondence, we constructed $3$-dimensional stationary black hole solutions that asymptote stretched AdS$_3$ space in Einstein gravity coupled to real scalar field matter. These solutions are similar to the so-called WAdS$_3$ black holes typically appearing in higher-curvature (massive) gravity in $3$-dimensions. However, here we study them as solutions of a parity-even field theory with second order field equations. The specific model we considered is the $3$-dimensional analogue of Horndeski theory; that is, the most general scalar-tensor theory yielding second order field equations. This means that, even when no-hair theorems in $3$-dimensions forbid the existence of this kind of solutions in the case of matter minimally coupled to gravity, the WAdS$_3$ black holes can still be supported by real scalar matter if the coupling with the curvature is of the Horndeski type. 

These black hole solutions have interesting properties; in particular, they have interesting thermodynamical properties such as finite Hawking temperature and entropy. We computed the entropy of the WAdS$_3$ black holes resorting to the Wald formula, and we speculated with the possibility of this entropy to admit a microscopic description in terms of a dual CFT$_2$. Assuming the existence of Virasoro symmetry in the dual theory led us to predict the value of the central charge (\ref{c}) of the dual theory via the Cardy formula.

We also studied other aspects related to holography, such as the inner black hole mechanics, whose statements led us to predict the values of conserved charges associated to the mass and the angular momentum of the black holes, given by (\ref{masssita}) and (\ref{angulito}) respectively. It would be interesting to confirm these values by an independent computation, and it would be also desirable to actually show the existence of Virasoro symmetry in the near boundary region. However, showing this would amount to first find a direct method to compute exact and asymptotic conserved charges around WAdS$_3$ space in the theory (\ref{action}). This is matter of future investigation\footnote{This program has been accomplished in other models; see for instance Ref. \cite{donnay2}}.

The results of this work can be thought of as an extension of the results of Ref. \cite{Bravo-Gaete:2014haa}, where the asymptotically AdS$_3$ black hole solutions in theory (\ref{action}) have been investigated. Here we have shown that WAdS$_3$ analogues to those solutions exist in the theory.

\[
\]

The authors thank Laura Donnay and Andr\'es Goya for discussions on related topics. The work of G.G. was partially funded by FNRS-Belgium (convention FRFC PDR T.1025.14 and convention IISN 4.4503.15), by the Communaut\'{e} Fran\c{c}aise de Belgique through the ARC program and by a donation from the Solvay family. It was also supported by grants PIP0595/13 and UBACyT 20020120100154BA, from Consejo Nacional de Investigaciones Cient\'{\i}ficas y T\'{e}cnicas\ and Universidad de Buenos Aires. The work of M.T. was partially supported by FONDECYT. The Centro de Estudios Cient\'{\i}ficos (CECs) is funded by the Chilean Government through the Centers of Excellence Base Financing Program of CONICYT- Chile.



\begin{thebibliography}{99}

\bibitem{Strominger} D. Anninos, W. Li, M. Padi, W. Song and A. Strominger, 
{JHEP} \textbf{0903} (2009) 130.

\bibitem{DHH} S. Detournay, T. Hartman and D. Hofman, Phys. Rev. D {\bf 86} (2012) 124018.



\bibitem{a1} G. Comp\`{e}re and S. Detournay, {JHEP} \textbf{0703}
(2007) 098.

\bibitem{a2} G. Comp\`{e}re and S. Detournay, {Class. Quant. Grav.} 
\textbf{26} (2009) 012001; Erratum-ibid. \textbf{26} (2009) 139801.

\bibitem{a3} G. Comp\`{e}re and S. Detournay, {JHEP} \textbf{0908}
(2009) 092.

\bibitem{a8} M. Blagojevi\'{c} and B. Cvetkovi\'{c}, {JHEP} 
\textbf{0905} (2009) 073.

\bibitem{a9} M. Blagojevi\'{c} and B. Cvetkovi\'{c}, {JHEP} 
\textbf{0909} (2009) 006.

\bibitem{a4} M. Henneaux, C. Mart\'{\i}nez and R. Troncoso, {Phys.
Rev.} \textbf{D84} (2011) 124016.

\bibitem{Geoffrey} G. Comp\`{e}re, M. Guica and M.J. Rodr\'{\i}guez, arXiv:1407.7871.

\bibitem{contextoa} M. Ba\~{n}ados, G. Barnich, G. Comp\`{e}re and A. Gomberoff, Phys. Rev. D {\bf 73} (2006) 044006.

\bibitem{a5} M. Guica,  JHEP {\bf 1212} (2012) 084.

\bibitem{a6} D. Anninos, M. Esole and M. Guica, {JHEP} 
\textbf{0910} (2009) 083.

\bibitem{anninos} D. Anninos, {JHEP} \textbf{0909} (2009) 075.

\bibitem{a7} D. Anninos, {JHEP} \textbf{1002} (2010) 046.


\bibitem{b1} W. Song and A. Strominger, {JHEP} \textbf{1205} (2012)
120.

\bibitem{b4} D. Israel, C. Kounnas, D. Orlando and P. Petropoulos, 
\textit{Fortsch. Phys.} \textbf{53} (2005) 73.

\bibitem{b5} S. Detournay, D. Orlando, P. Petropoulos and Ph. Spindel,%
{\ JHEP} \textbf{0507} (2005) 072

\bibitem{Israel:2003cx}
  D.~Israel,
and chronology protection,"
  JHEP {\bf 0401} (2004) 042.

\bibitem{b6} S. Detournay, D. Israel, J. M. Lapan and M. Romo, {JHEP} \textbf{1101} (2011) 030.

\bibitem{b7} T. Azeyanagi, D. M. Hofman, W. Song and A. Strominger, JHEP {\bf 1304} (2013) 078.

\bibitem{b8} G. Cl\'{e}ment, {Class. Quant. Grav.} \textbf{26}
(2009) 105015.

\bibitem{b9} G. Giribet and M. Leston, {JHEP} \textbf{1009}
(2010) 070.

\bibitem{c1} E. Tonni, {JHEP} \textbf{1008} (2010) 070.

\bibitem{c2} G. Giribet and A. Goya, JHEP {\bf 1303} (2013) 130.

\bibitem{c3} A. Goya, JHEP {\bf 1409} (2014) 132.

\bibitem{contextoz} O. Mi\v{s}kovi\'{c} and R. Olea, {JHEP} \textbf{0912}
(2009) 046.

\bibitem{corea} S. Nam, J. D. Park and S. H. Yi, Phys. Rev. D {\bf 82} (2010) 124049.

\bibitem{Celine} S. Detournay and C. Zwikel, JHEP {\bf 1505} (2015) 074. 

\bibitem{donnay} L. Donnay, J. Fern\'andez-Melgarejo, G. Giribet, A. Goya and E. Lavia, Phys. Rev. D {\bf 91} (2015) 125006.

\bibitem{donnay2} L. Donnay and G. Giribet, arXiv:1504.05640.

\bibitem{HofmanRolliest} 
  D.~M.~Hofman and B.~Rollier,
  Nucl.\ Phys.\ B {\bf 897} (2015) 1.

\bibitem{preKerrCFT} I. Bengtsson and P. Sandin, {Class. Quant. Grav.} 
\textbf{23} (2006) 971.

\bibitem{KerrCFT} M. Guica, T. Hartman, W. Song and A. Strominger, {Phys. Rev.} D \textbf{80} (2009) 124008.

\bibitem{Son}
 D.~T.~Son,
  Phys.\ Rev.\ D {\bf 78} (2008) 046003

\bibitem{BalasubramanianMcGreevy}
K.~Balasubramanian and J.~McGreevy,
  Phys.\ Rev.\ Lett.\  {\bf 101} (2008) 061601.

\bibitem{BTZ} M. Ba\~{n}ados, C. Teitelboim andJ. Zanelli, Phys. Rev. Lett. 
{\bf 69} (1992) 1849.


\bibitem{bha} K. Ait Moussa, G. Cl\'{e}ment and C. Leygnac, {
Class. Quant. Grav.} \textbf{20} (2003) L277.

\bibitem{bhz} A. Bouchareb and G. Cl\'{e}ment, {Class. Quant. Grav.} 
\textbf{24} (2007) 5581.


\bibitem{Horndeski:1974wa}
  G.~W.~Horndeski,
  Int.\ J.\ Theor.\ Phys.\  {\bf 10} (1974) 363.

\bibitem{Galileons}
A. Nicolis, R. Rattazzi and E. Trincherini, Phys. Rev. D {\bf 79} (2009) 064036.

\bibitem{Deffayet:2011gz}
  C.~Deffayet, X.~Gao, D.~A.~Steer and G.~Zahariade,
  Phys.\ Rev.\ D {\bf 84} (2011) 064039.

\bibitem{VanAcoleyen:2011mj}
  K.~Van Acoleyen and J.~Van Doorsselaere,
  Phys.\ Rev.\ D {\bf 83} (2011) 084025.

\bibitem{fab4}
  C.~Charmousis, E.~J.~Copeland, A.~Padilla and P.~M.~Saffin,
  Phys.\ Rev.\ D {\bf 85} (2012) 104040.
  C.~Charmousis, E.~J.~Copeland, A.~Padilla and P.~M.~Saffin,
  Phys.\ Rev.\ Lett.\  {\bf 108} (2012) 051101.

\bibitem{Charmousis:2014mia}
  C.~Charmousis,
  Lect.\ Notes Phys.\  {\bf 892} (2015) 25.

\bibitem{Kolyvaris:2011fk}
  T.~Kolyvaris, G.~Koutsoumbas, E.~Papantonopoulos and G.~Siopsis,
  Class.\ Quant.\ Grav.\  {\bf 29} (2012) 205011.

\bibitem{Rinaldi:2012vy}
  M.~Rinaldi,
  Phys.\ Rev.\ D {\bf 86} (2012) 084048.

\bibitem{Babichev:2013cya}
  E.~Babichev and C.~Charmousis,
  JHEP {\bf 1408} (2014) 106.

\bibitem{Anabalon:2013oea}
  A.~Anabal\'on, A.~Cisterna and J.~Oliva,
  Phys.\ Rev.\ D {\bf 89} (2014) 084050.


\bibitem{Minamitsuji:2013ura}
  M.~Minamitsuji,
  Phys.\ Rev.\ D {\bf 89} (2014) 064017.

\bibitem{Cisterna:2014nua}
  A.~Cisterna and C.~Erices,
  Phys.\ Rev.\ D {\bf 89} (2014) 084038.

\bibitem{Kobayashi:2014eva}
  T.~Kobayashi and N.~Tanahashi,
  PTEP {\bf 2014} (2014) 7,  073E02.

\bibitem{Graham:2014ina}
  A.~A.~H.~Graham and R.~Jha,
  Phys.\ Rev.\ D {\bf 90} (2014) 4,  041501.

\bibitem{Charmousis:2015aya}
  C.~Charmousis and D.~Iosifidis,
  J.\ Phys.\ Conf.\ Ser.\  {\bf 600} (2015) 1,  012003.
  
\bibitem{Babichev:2015rva}
  E.~Babichev, C.~Charmousis and M.~Hassa\"{\i}ne,
  arXiv:1503.02545.

\bibitem{Bravo-Gaete:2013dca}
  M.~Bravo-Gaete and M.~Hassa\"{\i}ne,
  Phys.\ Rev.\ D {\bf 89} (2014) 104028.

\bibitem{Charmousis:2014zaa}
  C.~Charmousis, T.~Kolyvaris, E.~Papantonopoulos and M.~Tsoukalas,
  JHEP {\bf 1407} (2014) 085.

\bibitem{Charmousis:2015txa}
  C.~Charmousis and M.~Tsoukalas,
  arXiv:1506.05014 [gr-qc].

\bibitem{Hui:2012qt}
  L.~Hui and A.~Nicolis,
  Phys.\ Rev.\ Lett.\  {\bf 110} (2013) 24,  241104.

\bibitem{Sotiriou:2013qea}
  T.~P.~Sotiriou and S.~Y.~Zhou,
  Phys.\ Rev.\ Lett.\  {\bf 112} (2014) 251102.





\bibitem{Bravo-Gaete:2014haa}
  M.~Bravo-Gaete and M.~Hassa\"{\i}ne,
  Phys.\ Rev.\ D {\bf 90} (2014) 024008.
  

\bibitem{Majo} A. Castro and M. Rodr\'{\i}guez, Phys.Rev. D {\bf 86} (2012) 024008.



\end{thebibliography}
\end{document}